\documentclass[10pt]{article}

\usepackage{amsmath}
\usepackage{amssymb}
\usepackage{enumerate}

\usepackage{graphicx}
\usepackage{algorithmicx}
\usepackage{algorithm}
\usepackage{subfigure}
\usepackage{cite}

\usepackage{color} 

\usepackage[labelfont=bf,labelsep=period,justification=raggedright]{caption}

\makeatletter
\renewcommand{\@biblabel}[1]{\quad#1.}
\makeatother

\date{}

\begin{document}
\begin{flushleft}
{\Large
\textbf{Fast $k$-NNG construction with GPU-based quick multi-select}
}
\\
Ivan Komarov$^{1}$, Ali Dashti$^{1}$,  
Roshan M. D'Souza$^{1\ast}$, 
\\
$^{\mathbf{1}}$ Dept. of Mechanical Engineering, Complex Systems Simulation Lab, University of Wisconsin-Milwaukee, Milwaukee, WI, USA
\\
$\ast$ E-mail: dsouza@uwm.edu
\end{flushleft}
\begin{abstract}
In this paper we describe a new brute force algorithm for building the $k$-Nearest Neighbor Graph ($k$-NNG). The $k$-NNG algorithm has many applications in areas such as machine learning, bio-informatics, and clustering analysis. While there are very efficient algorithms for data of low dimensions, for high dimensional data the brute force search is the best algorithm.  There are two main parts to the algorithm: the first part is finding the distances between the input vectors which may be formulated as a matrix multiplication problem. The second is the selection of the $k$-NNs for each of the query vectors. For the second part, we describe a novel graphics processing unit (GPU) -based multi-select algorithm based on quick sort.  Our optimization makes clever use of warp voting functions available on the latest GPUs along with use-controlled cache. Benchmarks show significant improvement over state-of-the-art implementations of the $k$-NN search on GPUs. 

\end{abstract}

\section{Introduction}
The $k$-nearest neigbhor ($k$-NN) and the related $k$-nearest neighbor graph ($k$-NNG) are important algorithms for data classification with a wide variety of applications in areas such as bioinformatics \cite{Roberts:2007, Weston:2004}, data mining\cite{Zaki:2000}, machine learning \cite{Maier:2009,Liu:2010,Tenenbaum:2000}, cluster analysis\cite{Franti:2006}, and pattern recognition\cite{Duda:2001}. Given a initial set of data points called corpus/training set, and a set of query points, $k$-NN find the nearest $k$ neighbors in the corpus set for each member of the query set. For low dimensional data-sets, there are a variety of indexing data structures such as  kd-trees \cite{Jones:2011}, BBD-tree \cite{Arya:1994}, random-projection trees (rp-trees) \cite{Dasgupta:2008}, and hashing based on locally sensitive hash \cite{Datar:2004} that can be built from the set of corpus/training data points. Next the $k$-NN search can be accomplished very efficiently for each data point in the query set using these data structures.  In the case of $k$-NNG, every data point in the corpus is also a member of the query set. \\

Direct construction of exact $k$-NNG has been also been investigated by many researchers \cite{Bentley:1980, Clarkson:1983,Vaidya:1989,ParedesCFN:2006,Chan:1998,ConnorK:2010}. All these methods have time complexity that increases exponentially with data dimension.  Approximate methods that can handle high-dimensional data with a trade-off between speed and accuracy have also been investigated. These techniques are based on hybrid of spatial subdivision and small scale brute force refinement  \cite{ChenFS:2009,Wang:2012, DongCL:2011}.  The spatial subdivision becomes grossly inaccurate for data sets that have dimension exceed $10^3$.\\

For high dimensional data, the most efficient method for finding $k$-NNG is in fact the brute force method \cite{Indyk:2004}. The brute force algorithm consists of two fundamental steps. The first step finds the distances between the query point based on an user-defined metric. This process result in a distance matrix of size $M \times N$ where $M$ is the number of query points and $N$ is the number of corpus data points. The $k$-NNs in the corpus data points for each query point is then found by sorting each row of the matrix and finding the $k$ indices of the $k$ smallest distances. Recent efforts have accelerated the brute force $k$-NN and $k$-NNG methods by parallelizing on graphics processing units. These methods differ in the manner in which the $k$ nearest indices are found. In \cite{Garcia2010, Arefin2012} each row of the distance matrix is processed by one thread. They use a modified insertion sort algorithm to select the $k$ nearest neighbors. The data structure is stored in global memory, which in turn leads to un-coalseced memory access patterns. Moreover, each thread has divergent paths to find the $k$ nearest neighbors which leads to loss in computational efficiency. In \cite{Barrientos2011}, every row in the distance matrix is processed by one thread block. A single heap in global memory maintains the current $k$ nearest elements. Each thread strides through the row of the matrix and inserts an element into a buffer if it is smaller that the largest element in the heap. When the buffer fills up, all threads synchronize and push their elements into the heap serially.   In \cite{Kato2009}, a thread block is used to process a single row. Each thread in the thread block strides through the array storing the $k$ smallest elements in a local heap maintained in global memory. Next all the thread heaps are merged into 32 heaps in shared memory threads in a single thread warp. Finally a single thread merges the 32 heaps in shared memory to find the $k$ nearest neighbors. Both methods dramatically slow down if $k$ grows large because of branch divergence and un-coalesced global memory access. In \cite{Kuang:2009}, radix sort is used for finding $k$ nearest neigbhors. Each row is processed in a separate kernel call and for $N$ that fit into GPU memory, this process underutilizes resources.\\

\subsection*{Graphics Processing Units: Architecture and Execution Model}
Graphics Processing Units (GPUs) were originally developed to handle computations related to computer graphics tasks such as transformation and shading. Computational scientists used this functionality to accelerate scientific computing by posing such computations in terms of shading operations \cite{Owens:2007}. Subsequently, GPU vendors developed specialized C language extensions to facilitate direct access to the computing hardware for scientific computing. Our implementation uses CUDA \cite{Sanders:2010}, a C language extension developed by NVIDIA.\\

Architecturally, a NVIDIA GPU consists of several multiprocessors (MPs). Each MP has several serial processors. In addition to registers, each MP has access to on-chip user-controlled cache called shared memory. Logically, threads are organized into thread blocks (TBs). All threads in a TB are executed on a single MP. Threads in a TB can be synchronized and can communicate through shared memory. Depending on the amount of shared memory and registers required, there could be several TBs assigned to be executed on a MP. All threads can communicate through global main memory. Registers have the highest access speed, followed by the shared memory, and then global memory. Of course, data in the registers cannot be shared among threads. Shared memory locations are accessed through specific memory banks.\\

At the hardware level, threads in a TB are divided into groups of 32 threads called thread warps. Thread warps are analogous the threads in a typical symmetric multi-processor architecture.  All threads in a thread warp execute in lockstep. Any branching between threads in a thread warp will cause serialization (different types of serialization depending on the complexity of the branching). Therefore it is advisable to avoid branching between threads of the same warp. During global memory access, if the threads in a warp access locations outside of contiguous 128 byte segments, the memory access will be serialized. Also, if two or more threads try to access shared memory affiliated to the same memory bank, this memory access will be serialized. An exception is when all threads in a warp access the same shared memory location. In such cases, a broadcast mechanism is automatically engaged. Therefore, implementations of algorithms for GPU acceleration require careful consideration of architectural limitations as well as advantages. \\

In our implementation we use an NVIDIA Tesla C2050 GPU. It has 14 multiprocessors each with 32 cores for a total of 448 cores. The single precision peak performance is 1.0TFlops. Each MP has 64KB of shared memory and a 128KB large register file. Main memory is 3GB GDDR5.

\section*{Methods}
In this paper, we present a hybrid parallelization approach for brute force computation of multiple $k$-NN queries. The first part of the brute force $k$-NN, namely, finding the distance metric matrix is formulated as a matrix multiplication problem. Therefore efficient implementations of dense matrix operations can be used. The second part is finding the $k$-NN for each query data point.  For this part of brute force $k$-NN, we have develop a two level parallel implementation of the quick select algorithm. At the top or coarse level, multiple queries are processed independently and in parallel. At the fine level of parallelism, every query is parallelized over thread in a thread block or thread warp. \\


\subsection*{Related Work}
Recently, there are quite a few works that are focussed on parallel $k$-NN queries on GPUs. In \cite{Sismanis:2012}, a truncated bitonic sort based selection algorithm has been described. The results show a dramatic drop off in performance with increase in both  the number of data points $n$ and number of nearest neighbors $k$. For example, the speedup over straight sort and selection using the CUDA \texttt{thurst::sort} for $n=2^{16}, k=2^8$, is barely 4X. In \cite{Baxter:2011}, a radix select implementation on the GPU is presented for selecting the $k^{th}$ element. This implementation is very efficient for large values of $n$ and runs upto 190X faster than a single core CPU implementation of the serial n-select algorithm. In \cite{Monroe:2011} a randomized selection algorithm is presented. This method however shows no improvement over CUDA \texttt{thurst::sort} for data sizes of upto $51e^5$. On the other hand the implementation is \cite{Baxter:2011} shows a 6.3X speedup. Both methods are however only suited for single query operations on ultra-large data sets. A similar implementation has been presented by  Alabi et al. \cite{Alabi:2012}.  The $k$-NN problem is slightly different from the method for selecting the $k^{th}$ element because extra work is necessary to track all the $k$ smallest elements. \\

In \cite{Cederman:2010} a data-parallel quick-sort algorithm for GPUs has been presented. They describe a two stage process. In the first stage, all thread-blocks cooperatively work on partitioning the original list to a point where the size of each partition is small enough to be processed by a single thread block. The first stage needs inter-thread block synchronization since the results of two or more thread blocks needs to be merged to build the partitions. This can either be done through atomic operations or through default synchronization by ending kernel execution. The latter needs a new kernel launch for each new phase of partitioning. The second stage is very similar to the first stage but each partition in this case is handled by a single thread block.\\

 The partitioning process is a two step processes and uses an auxiliary output array (i.e. it is not an in-place write operation). In the first step, a pivot element is randomly selected. Next, each thread strides through the input array counting the number of elements that are greater than equal to and less than the pivot. These two counts are stored in two arrays in shared memory. A pre-fix sum on these array in shared memory gives the location in global memory where each thread needs to write into the auxiliary the elements in the partitioning process. At the end of each partitioning task, the input and auxiliary arrays are swapped. When the partition size falls below a pre-defined threshold, a direct sort algorithm such as bitonic sort is used.\\

\subsection*{GPU-based quick multi-select}
Our approach while similar to approach in \cite{Cederman:2010}, address some of the major drawbacks as related to the $k$-NN application. For one, the quick-sort algorithm in \cite{Cederman:2010} is only suitable for sorting a single input array at a time. Secondly, inter thread block synchronization and synchronization between threads in the same thread block affects performance. Thirdly, the pre-fix sum operation that is used during the partition process is quite an expensive operation. Fourthly, the two step process of partitioning involves reading the same data from global memory twice (once for counting, once for writing). Finally, the write process into the auxiliary array is un-coalesced. Our approach operates on multiple arrays simultaneously. Each array is handled by a single thread warp. Threads in a warp are executed simultaneously on a single multi-processor and therefore are synchronization by default.  Partitioning of an array is done incrementally in 32 element wide segments. We use shared memory to ensure coalesced memory writes of the results of partitioning into the auxiliary array in global memory.  Our approach uses the warp voting function \texttt{\_ballot($p$)} to partition the input without reading the input array twice and without executing the parallel-prefix sum. The ballot function \texttt{\_ballot($p$)} fills a 32 bit unsigned integer, one bit per thread in the warp, based on the evaluation of the predicate $p$.\\
 
Each thread in the warp first reads in an element into its register from global memory. Elements are then written to a 32 element wide shared memory array with elements greater than equal to the pivot being written from the right end and elements less than the pivot being written from the left end. To do this, each thread needs to know where in shared memory to off load its element.  We execute the warp voting function based on a predicate which checks if the element in the register is greater than equal to the pivot or smaller than the pivot.  Threads which have an element greater than or equal to the pivot set the corresponding bit to '1' and to '0' if the element is less than the pivot (Figure 1). If the element in the register is less than the pivot, then the thread needs to find how many of threads before it have elements less than the pivot and vice-versa. We use a combination of bit shift operations and the \texttt{\_popc(x)} function on the integer result of the warp voting operation to accomplish this. The \texttt{\_popc(x)} functions counts the number of bits set to '1' in the input integer '$x$'. For example, if the warp vote integer in binary is $B=0101....$, then it is clear that threads 0,2  have elements less than the pivot and threads 1,3 have elements greater than equal to the pivot. Each thread $i$ within the warp computes the result $b=B>>(31-i)$.  When the  \texttt{\_popc($b$)} function is applied to the result of this step, it will indicate the position in shared memory at which thread $i$ will off load its element that happens to be less than the pivot. Similarly, $[31-\mbox{\texttt{\_popc}}(\tilde{b})]$ will indicate the position from the right side at which which thread $i$ will off load its element that happens to be greater than or equal to the pivot. In the example, thread 3 will bit shift $B$ to the right by 29 bits and the result will be $b=0101$. Then \texttt{\_popc($b$)=2}. Therefore thread 3 will off-load its element at the second location from the left in shared memory (Figure 2).  Note that the total number of elements in share memory that are less than the pivot is given by \texttt{\_popc($B$)}. Two global counters, $g_<$ and $g_{\ge}$, keep track of the total number of elements less than the pivot and total number of elements greater than equal to the pivot, respectively, are also maintained. These two counters indicate location in the auxiliary array at which the warp writes the incremental results of pivoting from shared memory.  Next the threads write the contents of the shared memory into global auxiliary array with threads whose id is less than \texttt{\_popc($B$)} (number of elements that are greater than the pivot) writing from the left side and other threads writing from the right side. This write process requires two coalesced writes, one for elements smaller than the pivot and one for elements greater than or equal to the pivot, into the auxiliary array (Figure 3). \\

With one warp per query, we must have a minimum of $1024$ queries to fully utilize GPU resources. If we use one thread block per query, we can reduce the minimum number of queries for full GPU utilization to $128$. When processing each row of the distance matrix using a thread block, each thread block will have several thread warps (depending on the thread count) that will cooperatively pivot elements. While writing the elements into shared memory, each thread has to determine its relative write position in shared memory w.r.t. all thread block threads. As a first step, each thread finds its write position w.r.t. threads in its own warp as described in the previous paragraph. Subsequently,  all threads in the thread block are synchronized. A inter warp pre-fix sum step determines the write position of each thread with respect to the entire thread block. The maximum number of thread warps per block is 16 (max. thread count in a thread block is 512 and the number of threads in a warp is 32). Because of the small number of elements, we use a single thread per thread block to conduct the inter warp prefix sum. This is much more economical than a multi-threaded pre-fix sum. \\

Once a partition is complete, the output array (auxiliary array) has a left side of length $L$ elements each of whom is less than the pivot. The right side is of length $R$ elements each of whom is greater than equal to the pivot. Suppose we need $K$ nearest neighbours, and $K<L$, then we need to process only the left hand side. Suppose $K>L$, then we keep the left hand side as is, and partition the right hand side to find $K-L$ elements. Since the input and auxiliary arrays are swapped at the end of the partition process, in the second case ($K>L$), we would have to copy the left hand side from the auxiliary array to the input array. We can avoid copying the data by instead storing a stack of references which indicate the start and end indices and the arrays (auxiliary or input) that the partitions that form the $K$ nearest neighbors are to be found. This significantly reduces the memory transactions needed for the operations.\\

\subsection*{Distance calculation}
Our implementation supports three distance metrics, namely, Euclidian, Cosine, and Pearson. The Euclidian distance between two vectors $\vec{x},\vec{y}$ is given by
\begin{equation*}
d_E(\vec{x},\vec{y}) = \sqrt{\Vert \vec{x} \Vert^2+\Vert \vec{y} \Vert^2-2\vec{x} \cdot \vec{y}} 
\end{equation*}
The Cosine distance metric is given by: 
\begin{equation*}
d_C(\vec{x},\vec{y}) = \frac{\vec{x} \cdot \vec{y}}{ \Vert \vec{x} \Vert \Vert \vec{y} \Vert}
\end{equation*}
Finally, the Pearson distance is given by:
\begin{equation*}
d_C(\vec{x},\vec{y}) = \frac{\hat{x} \cdot \hat{y}}{ \Vert \hat{x} \Vert\Vert \hat{y} \Vert}
\end{equation*}
where, $\hat{x} = \vec{x}-\bar{x}$ and $\bar{x}$ is the mean of the entries in $\vec{x}$. Note that the Pearson distance coefficient is essentially the Cosine distance of the centered data sets.\\

The process of finding either of these distances requires operations for finding vector means, vector magnitudes and dot products. For two data sets $X, Y$, where the columns of $X$ and $Y$ represent the data vectors $\vec{x}_i$ and $\vec{y}_j$ respectively, the $n\times m$ dot products 
\begin{equation*}
\vec{x}_i \cdot \vec{y}_j  \left \{ \begin{array}{c} i=1,2...n \\ j = 1,2..m \end{array} \right.
\end{equation*}
are easily formulated as a matrix product $X^TY$. There are very efficient libraries for dense matrix multiplication which can be used for this purpose \cite{Volkov:2008}. The process of finding vector means and vector magnitudes are essentially vector reductions. The \texttt{thrust} library has optimized \texttt{reduction\_by\_key} which we have used to simultaneously find the vector means and magnitudes of all vectors in $X,Y$.\\

Finding the Eulcidian and Cosine distances require the same inputs (dot products and norms of $\vec{x}, \vec{y}$). In the first parallel kernel operation, we compute the vector norms of the vectors in $X,Y$ using a combination of a transform iterator and \texttt{reduction\_by\_key} function. The transform iterator generates the square of individual elements and feeds the result into the \texttt{reduction\_by\_key} function which in turn computes the square of the vector norms.  The next step is to find the distance. In case of the Euclidian distance,  for finding the $k$ nearest vectors to $\vec{x}_i \in X$ from $\vec{y}_j \in Y$ the squared distance formula is given by:
\begin{eqnarray*}
d^2_{ij} = \Vert \vec{x}_i \Vert^2+\Vert \vec{y}_j \Vert^2-2\vec{x}_i \cdot \vec{y}_j
\end{eqnarray*} 
The comparison to find the $k$ nearest neighbhors of $\vec{x}_i$ is between the numbers $d^2_{ij}$. Furthermore, all of the $d^2_{ij}$ have a common addition factor $\Vert \vec{x}_i \Vert^2$. Therefore, it is sufficient to compare the distance metric $d'_{ij} = \Vert \vec{y}_j \Vert^2-2\vec{x}_i \cdot \vec{y}_j$. This saves an addition operation. For finding the $k$ nearest vectors to $\vec{y}_i \in Y$ from $\vec{X}_j \in X$,  we follow the same process but find the matrix of dot products $Y^TX$ by simply taking a transpose of the matrix $X^TY$ which we computed earlier. The same analysis is valid for Cosine distances as well.\\

For finding the Pearson distance, we need to pre-process the raw data to center it. We use the \texttt{reduction\_by\_key} to find the vector means.  We then center each vector using one thread block per vector. Once the data is centered, the process of finding the distance metric is the same as that of the Cosine distance. \\

\section*{Results}
Our algorithms were written in NVIDIA's CUDA \cite{Sanders:2010} and executed on a Tesla C2050 compute card. We used gcc4.2 with appropriate optimization flags. This card has 448 serial processors and a peak single precision performance of 1.03 TFlops.  The RAM memory is 3GB with a global memory bandwidth of 144 GB/s. We used CUDA 4.2 drivers for our implementation. We performed a comprehensive set of tests against implementations by Garcia et al.\cite{Garcia2010}, and Sismanis et al.\cite{Sismanis:2012}. Both these algorithms are for $k$-NN. We also benchmarked our algorithm against the  \texttt{nth\_element} from the standard template library and the GPU-based radix/bucket select implementation of $k^{th}$-element algorithm by Alabi et al.  \cite{Alabi:2012}. The work by Alabi et al., as indicated earlier, is slightly different since they just return the $k^{th}$ largest or smallest element and not $k$-NNs. This means that they do not keep track of the $k$-NN elements and therefore perform less work than our algorithm and the implementations by Garcia and Sismanis.  For accurate comparison, we downloaded the all GPU codes (Sismanis, Garcia, Alabi) and recompiled it using the same compiler and CUDA drivers as the ones we used to compile our code. The test data was a matrix ($Q \times n$, $Q$ is the number of queries, and $n$ is the size of the corpus data set) of uniformly random floating point numbers with indices. In some tests, indices were left out.\\

For large data sets, the code developed by Garcia et al. \cite{Garcia2010}., batches execution of multiple queries. For large $n$, the size of the batches (i.e., the number of queries) can cause underutilization of GPU resources. We chose parameters in our data sets $n,Q$ in a way that the algorithm by Garcia is operating at its full potential. Figure 4 shows the results of the benchmarks where kept $Q$ constant and varied the parameters $k,n$. We set the number of simultaneous queries $Q = 8192$. We varied the number of closest neighbours as $6 \le \log_2(k) \le 9$. Also, we varied the size of the corpus data as $13 \le \log_2(n) \le 18$. We only benchmarked the search part of the algorithm since the method for distance calculation is virtually identical between the two methods.  Figure 4 the performance advantage of our algorithm. There is dramatic gain in speedup when $k$ grow. As  $k$ grows, Garcia's algorithm shifts it's $k$-NN stack from shared memory to global memory and is saddled with all the associated pitfalls such as un-coalesced memory access.  In figure 5,  we illustrate the effect of changing the number of queries. In this example, we kept the size of the corpus at $n=65536$ and changed the number of queries as $8192 \le Q \le 65536$. Simultaneously, we changed the number of selections as $64 \le k \le 512$. For small queries, Garcia's method underutilizes the GPU.  As we increase the number of queries, the GPU utilization increases and therefore, we see a decrease in speedup from the peak. For the largest sizes of $k = 512$ and $Q = 65536$ that fit in our GPU memory, our method is $80 \times$ faster. \\

For the benchmarks against work by Sismanis (Figure 6), we compared the selection method alone. We kept the product $N \times Q = 2^{27}$ constant, where $N$ is the number data corpus elements (columns in the distance matrix) and $Q$ is the number of query elements (rows in the distance matrix). We plotted our speedup again the ratio $\log_2 \left(\frac{n}{Q}\right)$.  We ran a thousand trials for each evaluation and averaged the results. As seen in the figure 6, the performance gains for various $k$ rapidly rise with  $\log_2\left( \frac{n}{Q} \right)$, reaches a saturation level, and then falls for large values of  $\log_2\left( \frac{n}{Q} \right)$. The degradation with performance occurs at  $\log_2\left( \frac{n}{Q} \right)=13$. This is point corresponds to $Q=128$, i.e., where the number of thread blocks falls below $128$. At this stage, our algorithm is not able to fully utilize GPU resources. We show in a subsequent test that our algorithm saturates the GPU only when the thread block count exceeds $128$. The performance advantage grows with $k$. We were able to benchmark only upto $k=512$ as the Sismaniss implementation cannot handle $k>512$.  With increased GPU RAM, we can handle larger distance matrices and therefore, have more queries $Q$ for large values of $n$. For example, if we had a GPU with 18 GB of available RAM instead of the current 2GB, we would be easily able to handle distance matrices of sizes $128 \times 2^23$. This would push the point of performance degradation to a point given by $\log_2\left(\frac{n}{Q}\right) = 16$.  Consequently we envision our algorithm to perform even better on fused GPU-CPU architectures which have unified memory access to main CPU RAM. Such processors will become mainstream in the next 2-3 years.\\

Our third set of benchmarks were against the radix/bucket select-based $k^{th}$ statistic algorithm implemented by \cite{Alabi:2012}. This implementation takes one query at a time. Therefore, our comparison tests involved initially loading all the entire distance matrix into GPU RAM and then invoking the radix select algorithm one query at a time. This approach adds kernel launch overhead to the time for calculation. However, the typical kernel launch overhead for the GPU  in our tests is 5.7 micro seconds and therefore nearly two orders of magnitude smaller than the time required for task execution. Figure 7 shows the results of our benchmarks. For this battery of test, we kept the product $n \times Q = 2^{28}$ constant. Once again, we ran a thousand trials for each evaluation and averaged the results.  For low values of $\log_2\left(\frac{n}{Q}\right)$, our approach completely dominates. This is because in this regime, the implementation by \cite{Alabi:2012} underutilizes GPU resources. As the ratio $\log_2\left(\frac{n}{Q}\right)$ increases, our performance advantage decreases. At $\log_2\left(\frac{n}{Q}\right)=14$, our algorithm starts underutilizing GPU resources because the number of thread blocks in operation falls below $128$. The method of \cite{Alabi:2012} achieves GPU saturation only at $\log_2\left(\frac{n}{Q}\right)=18$ (i.e., $n=2^{23}, Q = 2^5$). Even at this point, our algorithm is slightly faster ($\approx 2 \times$), even though it is underutilizing GPU resources and is performing significantly larger amount of work keeping track of $k$-NNs instead of finding just the $k^{th}$ statistic.\\

Figure 8 illustrates our tests to investigate GPU saturation. In these tests we kept $n=2^{20}$ constant and varied $Q$, the number of queries from $Q=2^3$ to $Q=2^8$. For various values of $k$ it can be seen that the time per query drops with the $Q$ and flattens out at $Q=2^8$. Since we are utilizing one thread block per query, we should have at least $128$ queries to fully utilize GPU resources.  \\

Finally, we benchmarked our selection algorithm against the Standard Template Library \texttt{nth\_select}. Just as in \cite{Alabi:2012}, the \texttt{nth\_select} algorithm works on a single query. We compared our implementation with a single threaded \texttt{nth\_select} working one query at time. The data was loaded into memory ahead of time and queries were processed in a \texttt{for} loop.  Figure 9 shows the results of the benchmark. In this test we varied the parameter $k$ as $2\le k \le 1024$ and the parameter $n$ as $16 \le \log_2(n) \le 20$ .  The speedup increases with $n$ and as can be seen in the figure does not reach saturation even at $n=2^{20}$. There is slight degradation with increase in $k$ which is obvious since the larger $k$ entails keep track of a larger list of elements. In our tests, we achieved around $41 \times$ speedup for the largest $k$ with the largest data sets that could be accommodated in our GPU RAM. 

\section*{Discussion} We have successfully developed and implemented a multi-query $k$-NN algorithm on the GPU. Our algorithm is an implementation of the Quick Select method and uses the architectural advantages of modern GPUs. For corpus data that fit in GPU memory, based on our benchmarks, our algorithm outperforms all current state-of-the art methods. There are however many limitations to our method. To achieve full GPU saturation, we cannot have the number of queries $Q<128$,i.e., we need to have at least 128 thread blocks running simultaneously. Given the roughly 2GB available memory, this restricts the number of elements in the corpus data to $n<2097152$.  With the newer fused GPU-CPU processors, with unified RAM (upto 64 GB), we will be able to handle much larger data sets.  We are also looking at batch execution with data partitioning for large corpus data that do not fit into GPU memory. We may be able to overlap computation with data transfer to and from the CPU RAM to reduce overall execution time.  Batch execution will obviously require merging of results between executions and additional memory to store intermediate results. We believe there will significant opportunities for optimization based on the number of queries, corpus data size, and the dimension of vectors in the corpus data. This research will be explored in subsequent publications.

\section*{Figures}
\noindent {\bf  Figure 1: Read in process.}  Read in of the array is done incrementally in sets of 32 elements. As illustrated the memory access is coalesced. The value if stored in a register. Simultaneously, the invocation of the warp voting function fills the bit array $B$ based on the evaluation of the predicate which indicates if value in the register is greater than equal or less than the pivot.

\noindent{\bf Figure 2: Pivot process.} The pivot process is accomplished in shared memory. Each thread determines where in the shared memory the value has to be written. Values less than the pivot are accumulated on the left hand side and values greater than or equal to the pivot are accumulated on the right hand side. Since all threads write to different locations, there are no bank conflicts.

\noindent{\bf Figure 3: Write out process:} The thread id indicates (based on the computation \texttt{\_popc(B)} whether a given thread is writing out an element less than the pivot or greater than or equal to the pivot. The values $g_{< \mbox{ piv}}$ and $g_{\ge \mbox{ piv}}$ which are maintained in shared memory and updated incrementally indicate the location in the global array the location of the last element that is less than the pivot and greater than or equal to the pivot. This operation involves at most two coalesced memory writes.

\noindent{\bf Figure 4: Benchmark against GPU $k$-NN by Garcia  for selection alone with varying size of corpus data ($n$)\cite{Garcia2010}.}  We set the number of queries $Q=8192$. This ensured full GPU utilization. The parameter $n$ was varied from $2^{13}<n<2^{18}$. The number of nearest neighbors $k$ was varied as $64 \le k \le 512$. Except for $k=512$, the speedup remains flat w.r.t. $n$. The dramatic gain in performance with $k$ is because the Garcia algorithm shifts its $k$-NN stack from shared memory to global memory as $k$ increases. This increases inefficiencies due to uncoalesced memory access. We achieved a speedup $>350 \times$ for $k=512$ and $n=2^{18}$.

\noindent{\bf Figure 5: Benchmark of selection alone against GPU $k$-NN by Garcia for different query sizes ($Q$).} We set $n=65536$. We varied the number of queries as $2^10 \le Q \le 2^{13}$. Simultaneously, we varied the number of nearest neighbors as $64 \le k \le 512$.  Note that speedup falls as the number of queries $Q$ increases. This is because the Garcia algorithm increases GPU utilization as the number of queries is increased. For the largest data sets that fit into GPU memory, corresponding to the level of full GPU utilization by Garcia ($Q = 2^{13}$) and for the largest number of nearest neighbors ($k=512$), our algorithm is $80 \times$ faster.

\noindent{\bf Figure 6: Benchmark against work by Sismanis et al. \cite{Sismanis:2012}}. These tests were performed for elements that contained both values and indices. In this benchmark we varied $\log_2(n/Q)$ as $-3\le \log_2(n/Q) \le 20$. We kept the product $n\times Q = 2^{27}$. We simultaneously varied $k$ as $2\le k \le 512$. Performance peaks at  $\log_2(n/Q) = 7$ stays flat till  $\log_2(n/Q) = 11$ and starts falling at  $\log_2(n/Q) = 13$. The decrease in performance corresponds to the number of queries $Q$ falling below 128 where our method underutilizes GPU resources. 

\noindent{\bf Figure 7: Benchmark against work by Alabi et al. \cite{Alabi:2012}}. Here the benchmark is comparing the speedup of selecting $k^{th}$ element vs selection of $k$-NN using our algorithm. These tests were performed for elements with values alone. We kept the product $n\times Q = 2^{28}$. We simultaneously varied $k$ as $2\le k \le 1024$. Simultaneously, we varied $\log_2(N/Q)$ as $2 \le \log_2(N/Q) 22$. There is a dramatic fall in performance gain because the method by Alabi increases GPU utilization as $n/Q$ increases. It reaches saturation at  $n/Q = 2^{18}$ \cite{Alabi:2012}. Our algorithm starts underutilizing GPU resources at $n/Q = 14$. Even when Alabi et al. is saturated and our algorithm is significantly underutilizing GPU resources, we are $1.8\times$.  Once again, with a larger GPU RAM, our algorithm would perform significantly better.

\noindent{\bf Figure 8: Effect of varying the number of queries $Q$}. In this benchmark we kept $n=2^{20}$ and varied the number of queries as $8 \le Q \le 256$.  As can be seen from the graph for various values of $k$, the time per query decreases with increase in $Q$ and roughly flattens our at $Q=128$.

\noindent{\bf Figure 9: Benchmark against $n^{th}$ element}.  This benchmark is against a single core of the CPU. As can be seen, the performance degrades slightly with larger values of $k$. The number of queries were set to $Q = 256$. We ran up to $n=2^{20}$. We achieve around $41 \times$ speedup. As can be seen from the figure, the speedup graph has not saturated for the largest values of $n$ that can be accommodated on our GPU. 

\newpage
\begin{figure}[!h]
\label{readin}
\begin{center}
\includegraphics[scale=0.5]{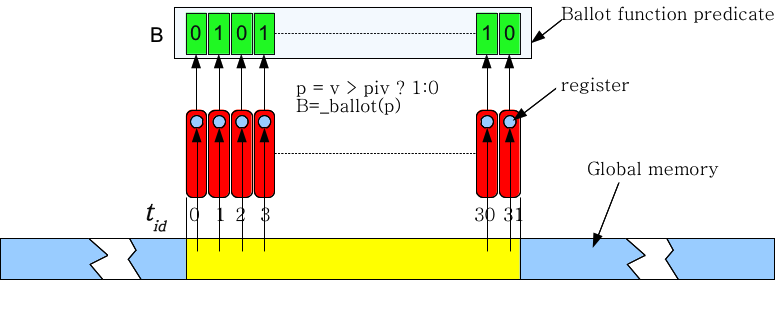}
\end{center}
\caption{}
\end{figure}

\newpage
\begin{figure}[!h]
\label{pivot}
\begin{center}
\includegraphics[scale=0.5]{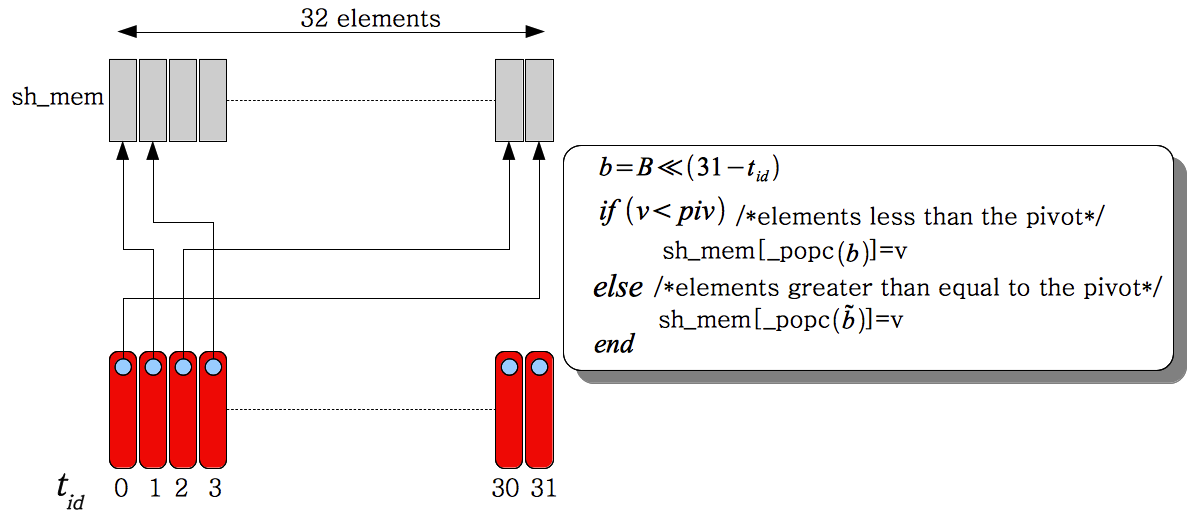}
\end{center}
\caption{}
\end{figure}

\newpage
\begin{figure}[!h]
\label{writeout}
\begin{center}
\includegraphics[scale=0.5]{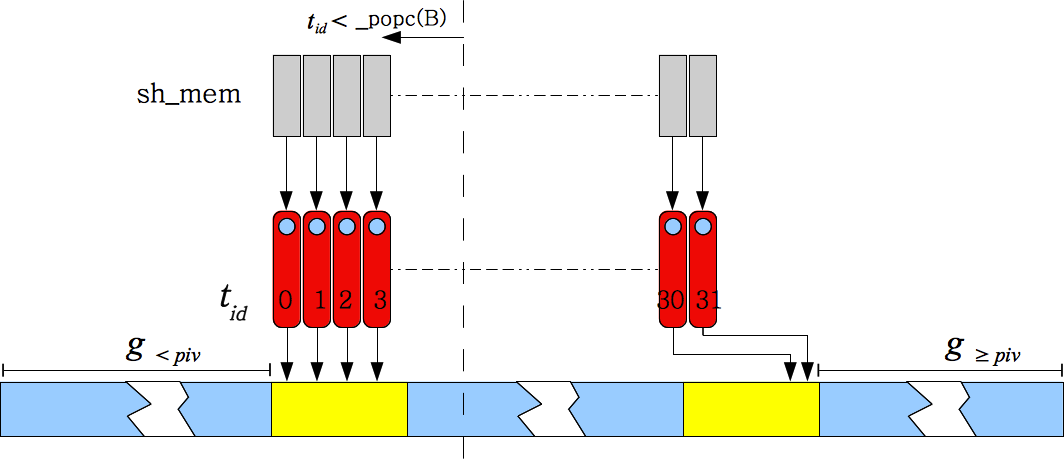}
\end{center}
\caption{}
\end{figure}

\newpage
\begin{figure}[!h]
\label{garcia}
\begin{center}
\includegraphics[scale=0.6]{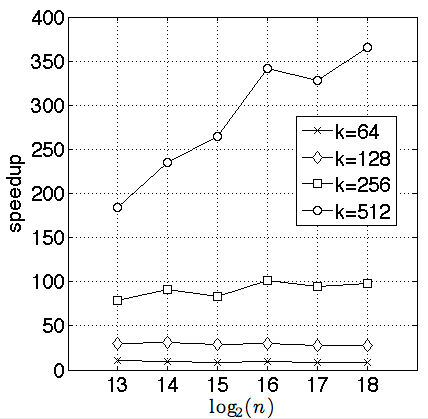}\\
\end{center}
\caption{}
\end{figure}

\newpage
\begin{figure}[!h]
\label{MGPU}
\begin{center}
\includegraphics[scale=0.6]{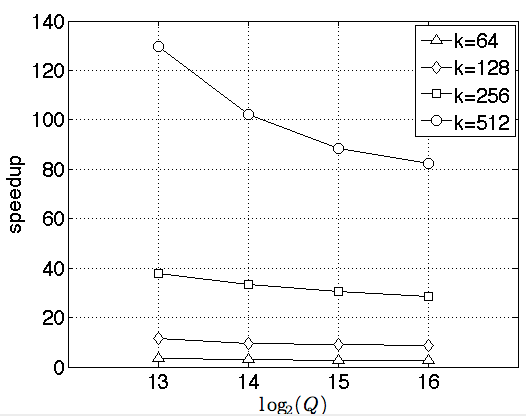}
\end{center}
\caption{}
\end{figure}

\newpage
\begin{figure}[!h]
\label{MGPU}
\begin{center}
\includegraphics[scale=0.6]{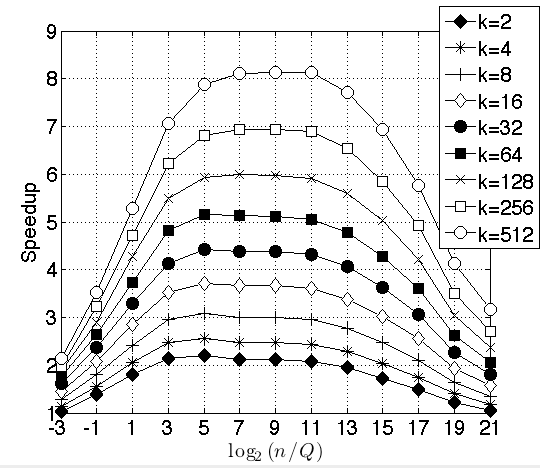}
\end{center}
\caption{}
\end{figure}

\newpage
\begin{figure}[!h]
\label{MGPU}
\begin{center}
\includegraphics[scale=0.6]{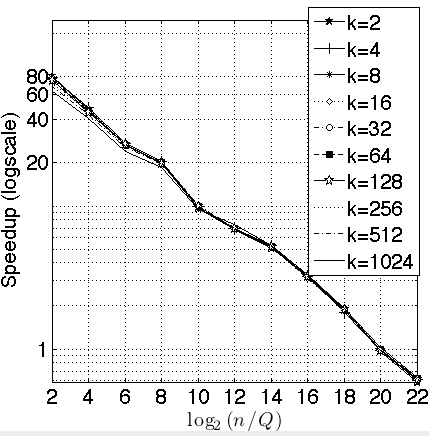}
\end{center}
\caption{}
\end{figure}

\newpage
\begin{figure}[!h]
\label{MGPU}
\begin{center}
\includegraphics[scale=0.6]{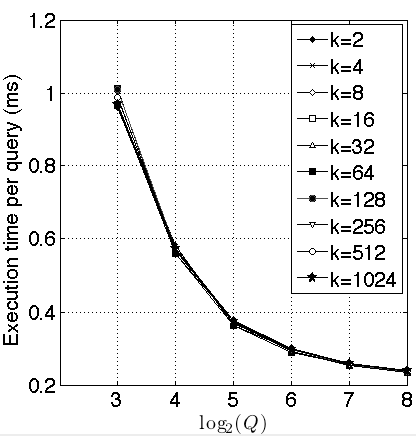}
\end{center}
\caption{}
\end{figure}

\newpage
\begin{figure}[!h]
\label{MGPU}
\begin{center}
\includegraphics[scale=0.6]{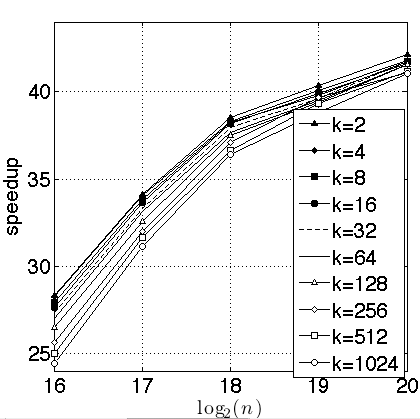}
\end{center}
\caption{}
\end{figure}

\bibliographystyle{plos2009}
\bibliography{GPU-kNN.bib}

\begin{thebibliography}{10}
\providecommand{\url}[1]{\texttt{#1}}
\providecommand{\urlprefix}{URL }
\expandafter\ifx\csname urlstyle\endcsname\relax
  \providecommand{\doi}[1]{doi:\discretionary{}{}{}#1}\else
  \providecommand{\doi}{doi:\discretionary{}{}{}\begingroup
  \urlstyle{rm}\Url}\fi
\providecommand{\bibAnnoteFile}[1]{%
  \IfFileExists{#1}{\begin{quotation}\noindent\textsc{Key:} #1\\
  \textsc{Annotation:}\ \input{#1}\end{quotation}}{}}
\providecommand{\bibAnnote}[2]{%
  \begin{quotation}\noindent\textsc{Key:} #1\\
  \textsc{Annotation:}\ #2\end{quotation}}
\providecommand{\eprint}[2][]{\url{#2}}

\bibitem{Roberts:2007}
Roberts A, McMillan L, Wang W, Parker J, Rusyn I, et~al. (2007) Inferring
  missing genotypes in large snp panels using fast nearest-neighbor searches
  over sliding windows.
\newblock Bioinformatics 23: i401--i407.
\bibAnnoteFile{Roberts:2007}

\bibitem{Weston:2004}
Weston J, Elisseeff A, Zhou D, Leslie CS, Noble WS (2004) {Protein ranking:
  from local to global structure in the protein similarity network.}
\newblock Proc Natl Acad Sci U S A 101: 6559--6563.
\bibAnnoteFile{Weston:2004}

\bibitem{Zaki:2000}
Zaki MJ, Ho CT, editors (2000) Large-Scale Parallel Data Mining, Workshop on
  Large-Scale Parallel KDD Systems, SIGKDD, August 15, 1999, San Diego, CA,
  USA, revised papers, volume 1759 of \emph{Lecture Notes in Computer Science}.
  Springer.
\bibAnnoteFile{Zaki:2000}

\bibitem{Maier:2009}
Maier M, Hein M, von Luxburg U (2009) Optimal construction of
  k-nearest-neighbor graphs for identifying noisy clusters.
\newblock Theor Comput Sci 410: 1749--1764.
\bibAnnoteFile{Maier:2009}

\bibitem{Liu:2010}
Liu W, He J, Chang SF (2010) Large graph construction for scalable
  semi-supervised learning.
\newblock In: ICML. pp. 679-686.
\bibAnnoteFile{Liu:2010}

\bibitem{Tenenbaum:2000}
Tenenbaum JB, Silva V, Langford JC (2000) {A Global Geometric Framework for
  Nonlinear Dimensionality Reduction}.
\newblock Science 290: 2319--2323.
\bibAnnoteFile{Tenenbaum:2000}

\bibitem{Franti:2006}
Fr{\"a}nti P, Virmajoki O, Hautam{\"a}ki V (2006) Fast agglomerative clustering
  using a k-nearest neighbor graph.
\newblock IEEE Trans Pattern Anal Mach Intell 28: 1875-1881.
\bibAnnoteFile{Franti:2006}

\bibitem{Duda:2001}
Duda RO, Hart PE, Stork DG (2001) {Pattern Classification (2nd Edition)}.
\newblock Wiley-Interscience, 2 edition.
\bibAnnoteFile{Duda:2001}

\bibitem{Jones:2011}
Jones PW, Osipov A, Rokhlin V (2011) {Randomized approximate nearest neighbors
  algorithm}.
\newblock Proceedings of the National Academy of Sciences 108: 15679--15686.
\bibAnnoteFile{Jones:2011}

\bibitem{Arya:1994}
Arya S, Mount DM, Netanyahu NS, Silverman R, Wu A (1994) An optimal algorithm
  for approximate nearest neighbor searching.
\newblock In: Proceedings of the fifth annual ACM-SIAM symposium on Discrete
  algorithms. Philadelphia, PA, USA: Society for Industrial and Applied
  Mathematics, SODA '94, pp. 573--582.
\newblock \urlprefix\url{http://dl.acm.org/citation.cfm?id=314464.314652}.
\bibAnnoteFile{Arya:1994}

\bibitem{Dasgupta:2008}
Dasgupta S, Freund Y (2008) Random projection trees and low dimensional
  manifolds.
\newblock In: Proceedings of the 40th annual ACM symposium on Theory of
  computing. New York, NY, USA: ACM, STOC '08, pp. 537--546.
\bibAnnoteFile{Dasgupta:2008}

\bibitem{Datar:2004}
Datar M, Immorlica N, Indyk P, Mirrokni VS (2004) Locality-sensitive hashing
  scheme based on p-stable distributions.
\newblock In: Proceedings of the twentieth annual symposium on Computational
  geometry. New York, NY, USA: ACM, SCG '04, pp. 253--262.
\bibAnnoteFile{Datar:2004}

\bibitem{Bentley:1980}
Bentley JL (1980) Multidimensional divide-and-conquer.
\newblock Commun ACM 23: 214-229.
\bibAnnoteFile{Bentley:1980}

\bibitem{Clarkson:1983}
Clarkson KL (1983) Fast algorithms for the all nearest neighbors problem.
\newblock In: FOCS. pp. 226-232.
\bibAnnoteFile{Clarkson:1983}

\bibitem{Vaidya:1989}
Vaidya PM (1989) {An O(n log n) algorithm for the all-nearest-neighbors
  problem}.
\newblock Discrete \& Computational Geometry 4: 101--115.
\bibAnnoteFile{Vaidya:1989}

\bibitem{ParedesCFN:2006}
Paredes R, Ch{\'a}vez E, Figueroa K, Navarro G (2006) Practical construction of
  {\it k}-nearest neighbor graphs in metric spaces.
\newblock In: WEA. pp. 85-97.
\bibAnnoteFile{ParedesCFN:2006}

\bibitem{Chan:1998}
Chan TM (1998) Approximate nearest neighbor queries revisited.
\newblock Discrete {\&} Computational Geometry 20: 359-373.
\bibAnnoteFile{Chan:1998}

\bibitem{ConnorK:2010}
Connor M, Kumar P (2010) Fast construction of k-nearest neighbor graphs for
  point clouds.
\newblock IEEE Trans Vis Comput Graph 16: 599-608.
\bibAnnoteFile{ConnorK:2010}

\bibitem{ChenFS:2009}
Chen J, ren Fang H, Saad Y (2009) Fast approximate {\it k}-nn graph
  construction for high dimensional data via recursive lanczos bisection.
\newblock Journal of Machine Learning Research 10: 1989-2012.
\bibAnnoteFile{ChenFS:2009}

\bibitem{Wang:2012}
Wang J, Wang J, Zeng G, Tu Z, Gan R, et~al. (2012) Scalable k-nn graph
  construction for visual descriptors.
\newblock In: CVPR. pp. 1106-1113.
\bibAnnoteFile{Wang:2012}

\bibitem{DongCL:2011}
Dong W, Moses C, Li K (2011) Efficient k-nearest neighbor graph construction
  for generic similarity measures.
\newblock In: Proceedings of the 20th international conference on World wide
  web. New York, NY, USA: ACM, WWW '11, pp. 577--586.
\newblock \doi{10.1145/1963405.1963487}.
\newblock \urlprefix\url{http://doi.acm.org/10.1145/1963405.1963487}.
\bibAnnoteFile{DongCL:2011}

\bibitem{Indyk:2004}
Indyk P (2004) Nearest neighbors in high-dimensional spaces.
\newblock In: Goodman JE, O'Rourke J, editors, Handbook of Discrete and
  Computational Geometry, Boca Raton, FL: CRC Press LLC. 2nd edition.
\bibAnnoteFile{Indyk:2004}

\bibitem{Garcia2010}
Garcia V, Debreuve E, Nielsen F, Barlaud M (2010) K-nearest neighbor search:
  Fast gpu-based implementations and application to high-dimensional feature
  matching.
\newblock In: ICIP. pp. 3757-3760.
\bibAnnoteFile{Garcia2010}

\bibitem{Arefin2012}
Arefin AS, Riveros C, Berretta R, Moscato P (2012) Gpu-fs-$k$nn: A software
  tool for fast and scalable $k$nn computation using gpus.
\newblock PLoS ONE 7: e44000.
\bibAnnoteFile{Arefin2012}

\bibitem{Barrientos2011}
Barrientos RJ, G\'{o}mez JI, Tenllado C, Prieto-Matias M, Marin M (2011) knn
  query processing in metric spaces using gpus.
\newblock In: Euro-Par (1)'11. pp. 380-392.
\bibAnnoteFile{Barrientos2011}

\bibitem{Kato2009}
Kato K, Hosino T (2009) Solving k-nearest vector problem on multiple graphics
  processors.
\newblock CoRR abs/0906.0231.
\bibAnnoteFile{Kato2009}

\bibitem{Kuang:2009}
Kuang Q, Zhao L (2009) A practical gpu based knn algorithm.
\newblock In: Proceedings of the Second Symposium International Computer
  Science and Computational Technology(ISCSCT ’09). pp. 151-155.
\bibAnnoteFile{Kuang:2009}

\bibitem{Owens:2007}
Owens JD, Luebke D, Govindaraju N, Harris M, Krüger J, et~al. (2007) A survey
  of general-purpose computation on graphics hardware.
\newblock Computer Graphics Forum 26: 80--113.
\bibAnnoteFile{Owens:2007}

\bibitem{Sanders:2010}
Sanders J, Kandrot E (2010) CUDA by Example: An Introduction to General-Purpose
  GPU Programming.
\newblock Addison-Wesley Professional, 1st edition.
\bibAnnoteFile{Sanders:2010}

\bibitem{Sismanis:2012}
Sismanis N, Pitsianis N, Sun X (2012) Parallel search of k-nearest neighbors
  with synchronous operations.
\newblock In: High Performance Extreme Computing (HPEC), 2012 IEEE Conference
  on. pp. 1-6.
\newblock \doi{10.1109/HPEC.2012.6408667}.
\bibAnnoteFile{Sismanis:2012}

\bibitem{Baxter:2011}
Baxter S (2011).
\newblock Mgpu select.
\newblock \urlprefix\url{http://www.moderngpu.com/select/mgpuselect.html}.
\bibAnnoteFile{Baxter:2011}

\bibitem{Monroe:2011}
Monroe L, Wendelberger J, Michalak S (2011) Randomized selection on the gpu.
\newblock In: Proceedings of the ACM SIGGRAPH Symposium on High Performance
  Graphics. New York, NY, USA: ACM, HPG '11, pp. 89--98.
\newblock \doi{10.1145/2018323.2018338}.
\newblock \urlprefix\url{http://doi.acm.org/10.1145/2018323.2018338}.
\bibAnnoteFile{Monroe:2011}

\bibitem{Alabi:2012}
Alabi T, Blanchard JD, Gordon B, Steinbach R (2012) Fast $k$-selection
  algorithms for graphics processing units.
\newblock J Exp Algorithmics 17: 4.2:4.1--4.2:4.29.
\bibAnnoteFile{Alabi:2012}

\bibitem{Cederman:2010}
Cederman D, Tsigas P (2010) Gpu-quicksort: A practical quicksort algorithm for
  graphics processors.
\newblock J Exp Algorithmics 14: 4:1.4--4:1.24.
\bibAnnoteFile{Cederman:2010}

\bibitem{Volkov:2008}
Volkov V, Demmel JW (2008) Benchmarking gpus to tune dense linear algebra.
\newblock In: Proceedings of the 2008 ACM/IEEE conference on Supercomputing.
  Piscataway, NJ, USA: IEEE Press, SC '08, pp. 31:1--31:11.
\newblock \urlprefix\url{http://dl.acm.org/citation.cfm?id=1413370.1413402}.
\bibAnnoteFile{Volkov:2008}

\end{thebibliography}

\end{document}